\documentstyle [12pt]{article}
\newcommand{\be}{\begin{equation}}
\newcommand{\ee}{\end{equation}}
\newcommand{\bear}{\begin{eqnarray}}
\newcommand{\eear}{\end{eqnarray}}
\newcommand{\lb}[1]{\label{#1}}
\newcommand{\bk}{\mbox{$k_{\bot}^{2}$}}

\newcommand{\F}{\mbox{$F(\omega,\mu^{2})$}}
\newcommand{\D}[2]{\mbox{${\cal D}_{#1}(#2)$}}
\newcommand{\Ll}[2]{\mbox{$L_{#1}^{#2}(\nu) $}}
\newcommand{\f}[2]{\mbox{$f_{#1}^{#2}$}}
\newcommand{\M}[2]{\mbox{$M_{#1}^{#2}$}}
\newcommand{\e}[1]{\mbox{Erfc$(#1)$}}

\newcommand{\al}{\alpha}
\newcommand{\bt}{\beta}

\newcommand{\de}{\delta}
\newcommand{\De}{\Delta}

\newcommand{\m}{\mu}
\newcommand{\ms}{\mu^{2}}
\newcommand{\n}{\nu}
\newcommand{\p}{\pi}
\newcommand{\ps}{\pi^{2}}

\newcommand{\la}{\lambda}
\newcommand{\om}{\omega}
\newcommand{\omm}{\omega_{max}}
\newcommand{\omr}{\omega_{run}}
\newcommand{\orr}{\omega_{R}}
\newcommand{\x}{\xi}

\newcommand{\op}{\omega^{+}}
\newcommand{\ops}{\omega^{+\, 2}}
\newcommand{\of}{\frac{\omega^{+}}{\omega}}
\newcommand{\as}{\alpha_{s}}
\newcommand{\gs}{g^{2}}
\newcommand{\Ns}{N^{2}}
\newcommand{\ov}{\frac{\omega}{\omega_{v}}}
\newcommand{\pams}{\frac{\partial}{\partial\ms}}
\newcommand{\dv}{\frac{d\om}{2\p i}}
\newcommand{\st}{\frac{-s}{t}}
\newcommand{\up}{u_{+}}
\newcommand{\um}{u_{-}}

\begin{document}
\pagenumbering{arabic}
\addtocounter{page}{0}

%%%SLAC-PUB Title Page%%%%%%%%%%%%%%%%%%%%%%%%%
\thispagestyle{empty}
\begin{flushright}
   \vbox{\baselineskip 12.5pt plus 1pt minus 1pt
         SLAC-PUB-6024 Rev.\\
         CU-TP-582     \\
         April 1993  \\
         (T)
             }
\end{flushright}

\begin{center}
{\bf High Energy Quark-Antiquark Elastic \\
  Scattering with Mesonic Exchange\footnote{Work supported
 in part by  Department of Energy
 contract DE-AC03-76SF00515.}}

Wai-Keung Tang \\
\vskip 1\baselineskip
Stanford Linear Accelerator Center \\
Stanford University, Stanford, CA 94309
\end{center}

\medskip

%\begin{center} {\bf Abstract} \end{center}

\begin{abstract}
 We studied the high energy elastic scattering of
 quark anti-quark with an exchange
of a mesonic state in the $t$ channel with $-t/\Lambda^{2}
\gg 1$. Both the normalization factor and the Regge trajectory can
be calculated in PQCD in cases of fixed (non-running)
and running coupling constant. The dependence of the
Regge trajectory  on
the coupling constant is highly non-linear and the trajectory is of
order of 0.2
in the interesting physical range.
\end{abstract}

%\vspace{1in}
\begin{center}
(Submitted to \it{Physical Review} \bf{D})
\end{center}
%%%%%BEGIN BODY OF PAPER%%%%%%%%%%%%%%%%%%%%%%%
\newpage
\section{Introduction}

With the advances of LHC and SSC, it is possible to study experimentally
the Regge behavior in the parton level where the momentum transfer
squared $-t \gg \Lambda^{2}$ but is still smaller than the center of mass
energy squared $s$
of the partons. The Regge limit of the parton
scattering amplitudes corresponds to the small $x$ limit of parton
distribution, while the presence of a ``large" scale $-t$ justifies
 the use of perturbative QCD.

The sea-quark and gluon distribution for small $x$ is related to the
Balitsky-Fadin-Kuraev-Lipatov
(BFKL) Pomeron~\cite{bd}. The distributions of sea-quarks and gluons
grow like $x^{-\al_{P}}$  at small $x$. where $\al_{P}$ is the trajectory of
the BFKL Pomeron. The behavior of the valence-quark distributions
is  controlled by the mesonic Reggeons \cite{bd}.
It grows as $x^{-\om_{R}}$
with $\om_{R}$ the trajectories of the mesonic Reggeons. However, great
care is necessary to separate the perturbative behavior from the
non-perturbative soft physics.

Several hard partonic processes  to measure the behavior
of the BFKL Pomeron [2-7] have been discussed in the
literature.
  In Ref \cite{tam},
A.H.~Mueller and the author proposed a process of high energy,
 fixed $t$ parton-parton scattering through the exchange of
a BFKL Pomeron. It is natural to extend this idea to the
process whereby a mesonic Reggeon is exchanged. It is the objective
of this paper to set up the necessary machinery to
investigate the process of the mesonic Reggeon exchange
and to
study the possibilty of measuring such behavior.
The scattering amplitudes are calculated, while the normalization
factor and the trajectory can be
obtained explicitly.

In this paper, we study the quark anti-quark scattering
amplitudes [Fig.1] with flavor exchange in the $t$ channel. In the
kinematic region,
\be
s \simeq \mid u \mid \gg -t \simeq \ms \gg \Lambda^{2},  \nonumber
\ee
where the auxiliary parameter $\mu$ is the infrared cutoff
of the transverse component of
 the momenta (with respect to the initial momenta $p_{1}$ and
$p_{2}$) of virtual particles in the Feynman integrals where the terms
$\sim \al(\mu^{2}) [(\al(\mu^{2})/\p) \ln^{2}(s/\ms)]^{n}$ are summed,
and $\alpha(\mu^{2})$ is the strong coupling constant at the scale $\mu^2$.
This is the double logarithmic (DL) approximation. The method of
separating the softest virtual particle \cite{ki} allows one to calculate the
partial wave amplitudes in the double logarithmic approximation. In Ref
\cite{li},
  R.~Kirschner and L.N.~Lipatov give equations
for the partial wave of the amplitudes  for both the color singlet and
the octet exchange in the $t$ channel. The octet exchange is suppressed,
because it has a strong tendency to radiate gluons in the rapidity
interval defined by the colliding quark anti-quark. This phenomenon
is reflected by the negative intercept of the
Regge trajectories of the octet
exchange. In what follows, we restrict ourselves to the consideration
of the color singlet exchange. Based on the partial wave results in Ref
\cite{li}, we study in detail the scattering amplitudes for  cases of
 both fixed (non-running) and running coupling constants, using analytic
methods as far as possible.

The outline of the paper is as follows: In Sec.~2, we review
the results of the partial wave amplitudes for the color singlet
exchange. Scattering amplitudes for the fixed coupling case are presented
in Sec.~3 for both positive and negative signature cases. In Sec.~4, both
approximate and numerical methods are employed to study the effect of
the inclusion of a running constant in the positive signature. Finally, Sec.~5
summarizes our conclusion.

\section{Review of the partial wave amplitudes}
We consider the amplitudes of annihilation ($q\overline{q} \rightarrow
Q \overline{Q}$) with the exchange of a mesonlike state in the $t$
channel. With respect to the color group $SU(N)$, the amplitudes can be
decomposed into a singlet $[M_{0}(s)]$ and an octet $[M_{v}(s)]$.
In double-logarithmic approximation, the spinor structure of the Born
term is preserved in higher order, so we can write
the amplitudes as
$b_{0}M(s,t)$ where $b_{0} =  \gamma_{\mu}^{\perp}
\otimes \gamma_{\mu}^{\perp}/s$ is the Born amplitude but without the
coupling constant $\gs$. $\M{0}{}$ and $\M{v}{}$ are
\bear
  \M{0}{} \mid_{Born} = \frac{\Ns-1}{2N}\gs &,& \M{v}{} \mid_{Born} = -
  \frac{1}{2N} \gs, \eear
after using the color projectors $P_{0}$ (singlet) and $P_{v}$
(octet)
\bear
  P_{o\; a b}^{\;\; a' b'} = \frac{1}{N} \de_{aa'}\de_{bb'} &,&
  P_{v\; a b}^{\;\; a' b'} =  \de_{ab}\de_{a'b'} - \frac{1}{N}
                                \de_{aa'}\de_{bb'}
\eear
and the decomposition
\bear
\M{ab}{a'b'}&=&P_{0\; a b}^{\;\; a'b'} \M{0}{} +
         P_{v\; a b}^{\;\; a'b'} \M{v}{}
\eear
where $a,\; b$ and $a', \; b'$ label the color states of the initial
and final quarks.

The asymptotics of the scattering amplitudes at large $s$ and fixed
$t$ are determined by the singularities of the partial wave $f_{j}(t)$
in the crossed channel. In order to express the amplitude in terms
of partial wave, we need to divide the amplitude $\M{0,v}{}$
into parts that are symmetrical and anti-symmetrical with respect
to the transformation $s\leftrightarrow u \simeq -s$ :
\bear
\M{}{\underline{+}}(s) & =& \frac{1}{2} [M(s) \underline{+} M(-s)].
\eear
In double-logarithmic approximation, the Sommerfeld-Watson
transformation reduces to the Mellin transformation, and the even (odd)
part of the amplitude is related to the positive (negative signature)
partial wave %
\bear
\M{}{p}(s/\ms) &=& \int_{a-i\infty}^{a+i\infty} \frac{d\om}{2\p i} \x^{p}
               (\om) \f{}{p}(\om) (\frac{s}{\m^{2}})^{\om},
\eear
with $\f{}{p}(\om)$ includes the factor $(sin\p \om)^{-1}$ usually
written explicity in the Sommerfeld-Watson integral. The signature
factor is given by
\bear
\x^{p}(\om) = &\frac{1}{2} (e^{-i\p \om} +p) & \simeq \left\{
              \begin{array}{ll}
                1 & p=+1 \\
                -\frac{1}{2} i \p \om & p=-1
              \end{array}
              \right. .
\eear

Using the method of isolating the softest virtual
particle with the lowest transverse momentum $\bk$ in the Feynman
diagrams, R.~Kirschner and L.N.~Lipatov were able to give
 equations for the partial wave amplitudes. The positive signature
amplitudes are
\bear
\f{0}{+}(\om) &=& \frac{a_{o} \gs}{\om} + \frac{1}{8\p^{2} \om}
                 (\f{0}{+}(\om))^{2}, {\rm and}            \nonumber \\
\f{v}{+}(\om) &=& \frac{a_{v} \gs}{\om} + \frac{b_{v}\gs}{8\p^{2} \om}
                  \frac{d}{d\om} \f{v}{+} (\om) + \frac{1}{8\p^{2} \om}
                 (\f{v}{+}(\om))^{2},
\eear
while the negative signature amplitude is more complicated,
\bear
\f{0}{-}(\om) &=& \frac{a_{0} \gs}{\om} - \frac{(\Ns-1)\gs}{4\p^{2} N\om}
                   \f{v}{+} (\om) + \frac{1}{8\p^{2} \om}
                 (\f{0}{-}(\om))^{2}
\eear
with
\bear
a_{0}=\frac{\Ns-1}{2N},  \;\;\;\;\;\;  a_{v} = -\frac{1}{2N}, \;\;\;\;\;
\;b_{v} = N   & &
\eear
and boundary conditions
%\
\bear
\f{i}{+} \mid_{\om \rightarrow \infty} &=& \frac{a_{i}\gs}{\om} \;\;\;
               \;\;                           (i=0,v).
\eear

Here we list the partial wave amplitudes that are relevant
to the color singlet exchange. The first terms in the above equations
are the contribution from the Born terms which have  pole singularity at
$\om = 0 $.

The equations of color singlet exchange are purely algebraic, so
 that the solutions can be written in explicit form:
\bear
\f{0}{+}(\om)
 &=& 4 \p^{2} \om \;
 \mbox{$[\; 1-\sqrt{1- (\frac{\op}{\om})^{2}}\;]$} \nonumber
\\
\f{0}{-}(\om) &=& 4 \p^{2} \om \;
\mbox{$[\; 1-\sqrt{1- (\frac{\op}{\om})^{2}
                           (1-\frac{1}{2\p^{2} \om} \f{v}{+})}\; ]$}
\lb{solf}
\eear
which shows that $\f{0}{+}(\om)$ has a square-root branch point at
(Fig.2)
\be
\om = \op = (\frac{\gs (\Ns -1)}{4 \ps N })^{1/2} =
(\frac{\as (\Ns-1)}{\p N})^{1/2},
\ee
while $\f{0}{-}(\om)$ has singularity to the right of $\op$ so that it
becomes dominant when energy is large. The equation for $\f{v}{+} (\om)$
is of Riccatti type. It can be solved easily and leads to the following
 result:
\bear
\f{v}{+}(\om) &=&  N\gs \frac{d}{d\om} \ln [ \exp (\frac{1}{4} (\ov)^{2} )
              \D{p_{v}}{\ov} ]  \nonumber \\
\om_{v}^{2} &=& \frac{\gs}{8\p^{2}} N, \;\;\;\;\; p_{v}= \frac{a_{v}}
                {b_{v}} = -\frac{1}{2\Ns},
\eear
where $\D{\n}{z}$ is the parabolic-cylinder function \cite{ma}.

In deriving  $\f{i}{}(\om)$,
the coupling constant was taken to be fixed.
But as the transverse momentum $\bk$ covers the large range from $\ms $
 to $s$ \cite{li},
 contribution from the running coupling is important and
 cannot be neglected. As demonstrated later, the inclusion of the running
coupling constant changes the singularity of the
partial
 wave amplitude from square root singularity  to pole singularity.

Taking into account the effect of running coupling, the
positive-signature singlet channel partial wave amplitude $\F$
satisfies
\bear
\ms \frac{\partial}{\partial \ms} \F - \om \F + \frac{1}{8\ps} (\F)^{2}
+ a_{0} \gs (\ms) &=& 0
\eear
with the boundary conditions
\bear
(\frac{\ms_{0}}{\ms})^{\om} \F \mid_{\ms \rightarrow \infty}
 &\rightarrow&0  \nonumber \\
\F \mid_{\om \rightarrow \infty } &\rightarrow&
\frac{a_{0} \gs (\ms)}{\om}.
\eear
The solution to the above equation is given by
\bear
\F &=& 8 \ps \ms \pams \ln [\Psi (-\frac{a_{0}}{8 \ps b \om}, 0;
         \frac{\om}{b\gs(\ms)} )]
\lb{solr}
\eear
where $b=(\frac{11}{3} N -2/3 N_{f})/16 \ps$, the coefficient of the
first term of  $\bt$ function. $\Psi(a,c;z)$ is a
confluent hypergeometric function \cite{ha} defined by
\bear
\Psi (a,c,z) &=& \frac{1}{\Gamma(a)} \int_{0}^{\infty} e^{-zt} t^{a-1}
(1+t)^{c-a-1} dt.
\eear
\section{Fixed coupling}
For the positive signature channel,
\bear
\M{0}{+}(s) &=& \int_{a-i\infty}^{a+i\infty}\dv (\st)^{\om} \f{0}{+}(\om)
\eear
where we have set $\ms=-t$. Substituting $\f{0}{+}(\om)$ from
eq.~(\ref{solf}), we have
\bear
\M{0}{+}(s)
 &=& \int_{a-i\infty}^{a+i\infty} \dv (\st)^{\om} 4 \ps \om \;
             \mbox{$[1-\sqrt{1-(\of)^{2}}]$},
\eear
where the contour of integration is shown in Fig.2.
We take the branch cut from $-\op$ to $+\op$, so the contour of
integration can be deformed to encircle the branch cut.
 The positive signature amplitude
becomes
\bear
\M{0}{+}(s)
&=& \int_{\op}^{-\op} \dv \;(\st)^{\om} \;4\ps \om \;\;
\mbox{Dis $[1-\sqrt{1-(\of)^{2}}]$ }.
\eear
With $\;\;\mbox{Dis $[1-\sqrt{1-(\of)^{2}}]$ }=2i
 \frac{\mid\ops-\om^{2}\mid^{1/2}}{\om} $ and modified Bessel function
 $I_{\n}$ defined by
\bear
I_{\n}(z) &=& \frac{(z/2)^{\n}}{\Gamma(\n+1/2)\Gamma(1/2)}
              \int_{-1}^{1} (1-t^{2})^{\n-1/2} e^{\underline{+}zt} dt.
\eear
$\M{0}{+}(s)$ is evaluated as
\bear
\M{0}{+}(s) &=& \frac{(2\p)^{2}\op}{y} I_{1}(\op y)
\eear
where $y=\ln(s/-t)$, the rapidity interval between the quark anti-quark
pair.
 If  $y$ is large, i.e. in the asymptotic region,
\bear
I_{1}(\op y) &\simeq& \frac{e^{\op y}}{\sqrt{2\p\op y}}    \nonumber
\eear
and
\bear
\M{asy}{+}(s) &=& \frac{(2\p)^{3/2} \ops}{(\op y)^{3/2}} e^{\op y}
\lb{mpasy}
\eear
which shows the Regge limit behavior.

As mentioned before, the negative signature channel dominates
asymptotically, so  an explicit solution is desired. However, because
 $\f{v}{+}$ depends on the parabolic-cylinder function, it is
not possible to have a convenient solution. In view of that, we
take lim $N \rightarrow \infty$. The  limit is not just an
academic exercise but has its own significance, because the
relevant parameter in this problem is $p_{v} =-1/2\Ns = -1/18$ which
 can be taken to be zero without introducing much  error.
With \cite{ma}
\bear
\frac{d}{dz} (e^{z^{2}/4} \D{\n}{z} ) &=& \n e^{z^{2}/4} \D{\n-1}{z},
\nonumber
\eear
$\f{v}{+}(\om)$ can be rewritten as
\bear
\f{v}{+}(\om) &=& -\frac{1}{2N} \frac{\gs}{\om_{v}}
\frac{\D{p_{v}-1}{\ov}}{\D{p_{v}}{\ov}}.
\eear
In the limit, $p_{v} = -1/2N^{2} \rightarrow 0$ while
\bear
\ops = \frac{\as (\Ns-1)}{\p N} \rightarrow \frac{\as N}{\p}
  = 2 \om_{v}^{2}.
\lb{omega}
\eear
 With the help of \cite{ma},
\bear
\D{0}{z} &=& e^{-z^{2}/4},~~ {\rm and}   \nonumber  \\
\D{-1}{z} &=& \sqrt{\frac{\p}{2}} \;e^{z^{2}/4} \;\e{\frac{z}{\sqrt{2}}},
\eear
where $\e{z}$ is the complementary error function.
$\f{v}{+}(\om)$ can be simplified as
\bear
\f{v}{+}(\om) &=& -\frac{2\ps \op }{\Ns} \;e^{(\of)^{2}} \;\e{\of}.
\eear

Define $u=\of$, and drop the first term of the negative partial wave
amplitude in eq.(~\ref{solf}) as it has no singularity and hence
 contributes nothing to the amplitudes after performing the Mellin
transformation. $\f{0}{-}(\om)$ then reduces to
\bear
\f{0}{-}(\om) &=& -\frac{(2\p)^{2}\op}{u} \sqrt{
             u[u^{3}-u-\frac{\p^{1/2}}{\Ns} \;e^{u^{2}} \e{u} ]}.
\eear
Expression (29) is corrected up to the order $1/{N^2}$ and clearly the
correction term comes from $\f{v}{+}(\om)$. The zeros of the expression
$u^{3}-u-\frac{\p^{1/2}}{\Ns} \;e^{u^{2}} \;\e{u}$  evaluated
numerically  are found to be $u_{+} = 1.0388$ and
$\um = -0.4668$. As $\up > 1$, the square root singularity of the
negative channel lies to the right of that of the positive signature
channel. But $\up-1 = .0388$  is not large. This indicates that
both channels have similar behavior phenomenologically.
The expression
$u^{3}-u-\frac{\p^{1/2}}{\Ns} e^{u^{2}} \e{u}$ can be approximated
by $(u-\up)(u-\um)^{2}$ within the region $\um\leq u\leq\up$ which
 is chosen to be the branch cut. With this approximation,
\bear
\f{0}{-}(\om) &=& -\frac{(2\p)^{2}\op}{u} (u-\um) \sqrt{
                   u(u-\up)}.
\eear

Performing the Mellin tranformation by deforming the contour of
integration to enclose the branch cut from $u=0$ to $u=\up$
(Fig. 3) leads to
\bear
\M{0}{-}(s) &=& \ps (\op)^{3} \int_{0}^{\up} du\; (\st)^{\op u}
            \;  (u-\um)\;\;\mbox{Dis} \sqrt{u(u-\up)}  \nonumber \\
 &=& 2i \ps (\op)^{3} \int_{0}^{\up} du \; (\st)^{\op u}
             \; (u-\um)\mid u(u-\up)\mid^{1/2}.
\eear
Change  variable  $u'~{\rm to}~u-\up/2$, and the integral can be
expressed in terms of the modified Bessel function $I_{\n}$,
\bear
\M{0}{-}(s) &=& i \frac{\p^{3} \ops}{y} \; e^{\up\op y/2}\;\;
              [(\frac{\up}{2} - \um-\frac{2}{\op y}) I_{1}(\up
                \op y/2) \nonumber \\
             & &   + \frac{\up}{2} I_{0} (\up \op y/2) ].
\eear
In the asymptotic region where $y$ is large, the amplitude
reduces to
\bear
\M{asy}{-}(s) &=&  \frac{i\sqrt{2}}{4}\p(\up-\um)\;\op \;e^{(\up-1)\op y}
              \;\;  \M{asy}{+}(s)
\eear
with $\M{asy}{+}(s)$ given by eq.~(\ref{mpasy}).
In the range of SSC energies
 where $y\sim 8$ and $\op \sim 0.5$,
\bear
  \frac{\sqrt{2}}{4}\p(\up-\um)\;\op \;e^{(\up-1)\op y}
&\sim& 1,
\eear
which suggests that
 the positive and negative signature amplitudes are of the
same order and cannot be distinguished in the interesting energies
range. Actually, $y$ needs to be about 50 before
the negative signature channel is
appreciatively different from the positive signature channel.
The difference between
$\sqrt{2} \om_{v}$ and $\op$ in eq.~(\ref{omega}) enhances
the negative signature channel a little bit. Effectively, it changes
$\up -1$ = 0.0388 to 0.1, but its effect  is still small at SSC
energies.

The two amplitudes have a phase difference of $\sim \p/2$, because
$\M{asy}{+}(s)$ is purely real while $\M{asy}{-}(s)$ is purely imaginary,
and they have nearly the same magnitude.
Both  give a non-linear quark anti-quark Regge trajectory
$\om(t) \sim \sqrt{\as(-t)}$.

\section{Running Coupling}
  As we have already mentioned, there is reason to believe that
the inclusion of running coupling effects will greatly change the
behavior of the amplitudes. In this section, we would like to
restrict ourselves to the study of
the positive signature singlet channel.
 It is plausible to suggest that
the inclusion of the running coupling effect would not change
the similarity between the positive and negative signature
 amplitudes greatly for the following reasons: the only
difference between the positive and negative channels stems from
the contribution of the double logarithmic soft gluon;
the soft gluon
 contributes to the negative signature singlet channel but not
the positive signature singlet channel~\cite{li}.
 However, as seen
in the previous section, the soft gluon does not change
the amplitude appreciatively. The inclusion of the running coupling
will decrease the importance of the soft gluon contribution as the
coupling between quark and gluon is smaller than in the
fixed coupling case. Therefore, for practical purposes, we can
 take positive and negative singlet amplitudes
 to be equal in magnitude but with phase difference $\p/2$.

In the following subsections, we will
 study the effect of the running coupling in the positive
signature channel using two different methods. In the first
method, we take lim  $b \rightarrow 0$, where $b$ is the coefficient
of the first term in the $\bt$ function. In this
approximation, we will recover the previous $\f{0}{+}(\om)$
result with a correction term which is liner in $b$. However,
as we will show later, the correction term enhances the amplitude
by $(\ln y)^{3/2}$ relative to the $\f{0}{+}(\om)$ term, so that the
correction term dominates asymptotically and the approximation
breaks down. The breakdown of $b$ expansion in the asymptotic region
closely relates to the fact that the transverse momentum in
 the loop integral extends to the order of $s$. It illustrates
that in the asymptotic region, the running coupling effect
 is not a small perturbation to the fixed coupling, but rather,
it changes the behavior of the amplitudes dramaically.

In the second method, we take the approximation $\la = a_{0}/
(8\ps b\om) \rightarrow \infty$. It gives a fairly accurate
result with $\sim 10\%$
error  compared to the numerical calculation, even though
the value of $\la$ lies between 1 and 2 when the running coupling
constant is in the interesting physical range.

Let us define
\bear
\la = \frac{a_{0}}{8\ps b \om}, \;\;\;\;\;& &
\n = \frac{\om}{\gs(\ms) b}.
\eear
Both $\la$ and $\n$ are positive if $\om > 0$. In the color
singlet channel, the trajectories lie to the right of the
origin of the complex $\om$ plane, because the higher order diagrams
enhance the amplitudes. This is in contrast to the case of the color octet
exchange which suppresses the amplitude. It is safe
to assume that both $\la$ and $\n$ are positive. As $\Psi$'s  dependence
on $\ms$ is through the running coupling constant $\gs(\ms)$,
it is better to write
\bear
\pams  = \ms \frac{\partial\gs}{\partial \ms} = \om \frac
{\partial}{\partial \n} & &
\eear
where $
 \ms \frac{\partial\gs}{\partial \ms} = - b g^{4} $.
In terms of $\la$ and $\n$, the partial wave amplitude $\F$
is
\bear
\F &=& 8\ps \om
\frac{\partial}{\partial \n} \ln [ \Psi(-\la,0;\n) ].
\lb{msol}
\eear
\subsection{$b$ expansion}
In the limit $b \rightarrow 0$, $\la, \n \rightarrow \infty$
; $\Psi$ can be approximated by \cite{ha}
\bear
\Psi(-\la,0;\n) &=& \la^{\la} \n^{1/4} (\n -4 \la)^{-1/4}  \nonumber \\
& &
\exp(-\la+\frac{\n}{2}-\frac{1}{2} \n^{1/2} (\n-4\la)^{1/2}
+\la \ln\mbox{$[\frac{(\n^{1/2}+(\n-4\la)^{1/2})^{2}}{4\la}]$})
\eear
which leads to
\bear
\frac{\partial}{\partial \n} \ln [ \Psi(-\la,0;\n) ]
&=& \frac{1}{2} [ 1-(1-\frac{4\la}{\n})^{1/2} -
\frac{2\la}{\n(\n -4\la)} ].
\lb{bexp}
\eear
The second term in the above expression turns out to be independent
of $b$ and with the first term, it
recovers $\f{0}{+}(\om)$. The third term,
which is the correction term, is found to be
\bear
\frac{2\la}{\n(\n -4\la)} &=&
 \frac{2\p b \as \ops}{\om (\om^{2} -\ops)}.
\lb{third}
\eear
It is linear in $b$ and has a pole singularity at $\om=\op$ in
contrast to the square root singuarity of the first term. Putting
eqs.~(\ref{msol}), (\ref{bexp}), and (\ref{third}) together,
we get
\bear
F(\om) &=& \f{0}{+}(\om) -
 \frac{(2\p)^{3} b \as \ops}{\om^{2} -\ops}.
\eear
The negative sign of the correction terms indicates that the
running coupling reduces the overall amplitude. This behavior
is also found in the Pomeron exchange \cite{kw}. The contribution of
the correction term to the positive signature amplitude
is
\bear
 \De \M{0}{+} (s) &=& -\frac{(2\p)^{3}}{2} b \as \op \; e^{\op y}
\eear
which leads, in the asymptotic region, to
\bear
\M{run}{+} &=& \M{0}{+} + \De \M{0}{+} \nonumber \\
    &=& [ 1-(2\p)^{1/2} \p b \as \op y^{3/2} ] \M{asy}{+}
\eear
where the correction term has $y^{3/2}$ dependence relative to the
first order term. The approximation is valid only when
\bear
(2\p)^{1/2} \p b \as \op y^{3/2} \ll 1.     & & \nonumber
\eear
With $\op \sim 0.4$, $\as \sim 0.20$ and $b = 0.053$ (four flavors),
the above inequality implies $y \ll 9.7$ which is  the boundary of
the asymptotic region.

\subsection{$\la \rightarrow \infty$ approximation}
Before taking any approximation, let us study the nature of the
singularity of the partial wave amplitude $\F$,  given
by eq.~(\ref{msol}). Both  $\la$ and $\n$ are positive
provided that $\om$, the Regge trajectory, is positive. Using
the differential property of the confluent hypergeometric
function \cite{ha},
\bear
\frac{\partial}{\partial \n} \Psi(a,c,;\n) &=& -a \Psi (a,c+1;\n)
\eear
$\F$ becomes
\bear
\F&=& 8\ps \om \frac{\la \Psi(1-\la,1;\n)}{\Psi(-\la,0;\n)}
\label{mmsol}
\eear
$\Psi(a,c;\n)$ is a many-valued function of $\n$, and we usually take
its principal branch in the plane cut along the negative real
axis. Therefore, $\Psi(a,c;\n)$ is analytic for $\n > 0$. For
$\n >0 $, the singularity of $\F$ must be the zeros of $\Psi(-\la,0;\n)$.
 $\Psi(a,c;\n)$ cannot have positive zeros if $a >0$ or $1+a-c>0$
\cite{ha}.
This implies that $1<\la$ must be true for $\Psi(-\la,0;\n)$ to have positive
zeros. From the definition of $\la$, $\la \sim 1/\om$, where $\om_{max}$
corresponds to $\la =1$, and $\Psi$  has one zero. Actually,
according to Ref~\cite{ha},  the range  $1\leq\la<2$, $\Psi (-\la
,0;\n) $ has one zero. This can be seen by noting that \cite{ha}
\bear
\Psi(-n,c;\n)&=&n! (-1)^{n} \Ll{n}{c-1}
\eear
where $\Ll{n}{\mu}$ is the generalized Laguerre polynomial \cite{ma}.
Therefore
\bear
\Psi(-1,0;\n) & =& (-1) \Ll{1}{-1} \nonumber \\
 &=& \n
\eear
as $\Ll{n}{-n} = (-\n)^{n}/n!$. The zero of $\Psi(-1,0;\n)$ is at $\n=0$
which in turn gives $\as \rightarrow \infty$ as $\om=\om_{max}=
a_{0}/(8\ps b) =0.32$ is fixed by $\la=1$. For $\la =2$,
which means $\om=\om_{max}/2=0.16$,
\bear
\Psi(-2,0;\n) &=& -\n (2-\n)
\eear
where $\Ll{2}{-1} = -\n(2-\n)/2$ is used. Both $\n=0$ and $\n=2$
are the zeros of $\Psi(-2,0;\n)$, $\n =2$ corresponding to $\as
=\om_{max}/(16\p b)=0.12$. The $\n=0$ solution is the subleading
trajectory and so we do not consider here.
 Hence, from $\la=1$ to
$\la=2$, $\as$ ranges from 0.12 to $\infty$ which covers a large
range of
the interesting physical region.

{}From the above discussion, we see that interesting physics already
lies inside the range $1\leq \la \leq 2$. $\la $ is of order 1,
so it is not very large. Neverthless, let us take a large $\la$
limit and compare the $\om(\as)$ obtained
with the exact results at $\as=\infty$ and $\as=0.12$.

When $\la\rightarrow \infty $, $\Psi(-\la,0;\n)$ can be approximated
as \cite{ha}
\bear
\Psi(-\la,0;\n) &\simeq& 2^{1/2} \la^{\la-1/4} \n^{-1/4}
                     \; e^{\n/2-\la} \cos(\la \p - 2(\la\n)^{1/2}
                   -\frac{\p}{4}),
\eear
which in turn gives
\bear
\F &=& 8\ps \om [-\frac{1}{4\n} +\frac{1}{2} + (\frac{\la}{\n})^{1/2}
      \tan ( \la \p -2(\la\n)^{1/2} -\frac{1}{4}) ].
\label{laexp}
\eear
The first two terms can be dropped as they contribute nothing to the
scattering amplitude after performing the Mellin transformation.
{}From Eq.~(\ref{laexp}), we see that $\F$ has pole singularities at
\bear
\la \p - 2(\la\n)^{1/2} -\frac{1}{4}\p &=&  \frac{n}{2}\p,
\eear
where the leading trajectory corresponds to  $n=1$,
\bear
\om_{run}(\as) &=& \frac{\om_{max}}{[\frac{3}{4} +
\frac{2}{\p}(\frac{\om_{max}}{4\p b\as})^{1/2}]} \nonumber \\
               &=& \frac{\op}{[\frac{4}{\p} +\frac{6\pi^{2}b}{a_{0}}\op]}.
\label{regge}
\eear
When $\as=\infty$, $\om_{run}=4\omm/3$ which is $33\%$ larger than the
exact value. For $\al = \omm/(16\p b) =0.12$, $\omr=0.99 \omm/2$,
which is only $1\%$ off.   The above analysis indicates that the $\omr$
obtained from the large $\la$ approximation is accurate in the region
 where the running coupling is small. Equation~(\ref{regge}) shows that
the trajectory of the mesonic Reggeon
is highly non-linear in $\as$, and one can
 see that $\omr(\as)$ is always less than $\op(\as)$.
The singularity of the partial wave amplitude in
 running coupling moves to the left  of that of
fixed coupling. For $\as=0.20$, a typical value, $\omr=0.19$
which is less than half of $\op=0.40$. The smallness of the
mesonic Reggeon trajectory imposes a serious problem of observing
the mesonic exchange in the parton level experimentally, especially
given that the mesonic exchange amplitude is already suppressed
by a factor of $s$ relative to
the Pomeron exchange which constitutes a strong background.

The amplitude can be obtained by the Mellin transforming
 Eq.~(\ref{laexp}),
\bear
\M{run}{+}(s) &=& 64\p^{3} \omega^2_{run} b (\frac{\as}{2\p
a_{0}})^{1/2} \; e^{\omr y}.
\label{amp}
\eear
There is no justification for the correctness of the
above expression
in large $\la$ approximation, although the trajectory does
give fairly accurate results. Therefore, it is natural to consider
the numerical evaluation of the amplitude and to compare it with the
approximated result obtained in this section. To our surprise,
the formula~(\ref{amp}) is accurate
 to within $10\%$,  as will be shown in the next section.
\subsection{Numerical Calculation}
Let us begin the numerical calculation by performing the Mellin
transformation on $\F$ using the expression (\ref{mmsol}). This
leads to
\bear
\M{run}{+} &=& \int \dv \;(\st)^{\om} \;8\ps \om\;
\frac{\la \Psi(1-\la,1;\n)}{\Psi(-\la,0;\n)} \nonumber \\
&=& 8\ps \om \;(\st)^{\om} \;n(\as) \mid_{\om=\orr}
\eear
with
\bear
n(\as)&=&  \frac{\gs b}{
    1+ \frac{\gs b}{\om \Psi(1-\la,1;\n)}\;
    \frac{d \Psi(\la,0;\n)}{d(-\la)}} \left.
    \right|_{\om=\orr} \nonumber \\
\eear
where $\orr$, the leading trajectory, is the root of $\Psi(-\la,0;\n)$. We are
interested in
the range $1\leq \la \leq 2$. The roots $\orr(\as)$ in this
range are shown in Fig.~4.
The large $\la$ approximation suggests that we use the following
parameterization to fit the data in Fig.4,
 where $\al$ and $\bt$ are  constants:
\bear
\orr(\as) &=& \frac{\omm}{\al + \bt \as^{-1/2}}
\label{tra}
\eear
Numerical fitting gives
\bear
\al =1.14 \times \frac{3}{4},~{\rm and} \;\;\;& & \bt=0.90 \times
         \frac{2}{\pi}(\frac{\omm}{4\pi b})^{1/2}
\eear
when we use the data from $1.5\leq \la \leq 2$  corresponding
to $0.121\leq \as \leq 0.382$.
Compared with $\al = 3/4$ and
$\bt= \frac{2}{\p} (\frac{\omm}{4\pi b})^{1/2}$ in
large $\la $ approximation, it once again confirms that it
is a good approximation for the trajectory $\orr$.

For the normalization $n(\as)$, the results for
$\as = 0.121$ to 0.382 are shown in Fig.~5. They can be summarized by the
following
 formula:
\bear
n(\as) &=& 0.895 8\p b \,\orr (\frac{\as}{2\p a_{0}})^{1/2},
\eear
where $\orr(\as)$ is given by Eq.~(\ref{tra}). Therefore,
\bear
\M{run}{+}(s) &=& 0.895\times 64\p^{3} b \, \orr^{2}
                                 (\frac{\as}{2\p a_{0}})^{1/2}
                  e^{\orr y}
\eear
is our final result for the positive signature mesonic singlet exchange
 in the $t$ channel. The factor 0.895 indicates that the large
$\la$ approximation has a $10 \%$ error in the
normalization $n(\as)$.

%The phenomenological and physical consequences of the above analysis
%are considered in  a forthcoming paper \cite{come}.
%
\section{Conclusion}
In this paper  we analyze the scattering
amplitudes of quarks and anti-quarks through flavour exchange
in the Regge limit. Both the normalization factors and the
mesonic Reggeon trajectory are obtained.

The negative and positive signatures of the fixed coupling constant
case give nearly the same energy behavior at the SSC energies.
The Regge trajectories are porportional to $\sqrt{\al_{s}(-t)}$ which
are non-linear in strong coupling constant.

The inclusion of the effect of the running coupling
constant reduces the magntiude of the amplitudes dramatically.
The failure of small $b$ expansion indicates that the running coupling
effect is not a small perturbation relative to the fixed coupling
case. The change of the nature of singularty, from  a square
root branch cut to a  simple pole, reflects the importance of the
effect of the running coupling constant. Moreover, the position of
the singularity shifts to the left and is a lot smaller than that
of the fixed coupling case. Its smallness imposes a serious challange
to observing experimentally the mesonic exchange in the parton level.

Although in this paper we  study just the positive signature in the
running coupling case, as explained in Sec.~4, we do expect
that positive and negative signature amplitudes have approximated
normalization factors and trajectories. This enables us to make
an estimation on
the possibility of observing mesonic exchange in the parton level.
%The results will be reported in a forthcoming paper \cite{come}.
%

\begin{flushleft}{\bf Acknowledgements}\end{flushleft}

I am very much indebted to Professor A.H.Mueller for
suggesting  this work and for many stimulating discussions.
I am also indebted to Stanley Brodsky for constructive comments
on an earlier draft of this paper.

\newpage
\begin{flushleft}
{\bf Figure Captions}
\end{flushleft}
\begin{description}
 \item [Fig. 1:]  Elastic quark anti-Quark annihilation.
 \item [Fig. 2:]  The square-root branch cut of $\f{0}{+}$. $C$ is
 the contour of the Mellin transformation  which lies to the right
 of the singularities.
 \item [Fig. 3:]  The square-root branch cut of $\f{0}{-}$.
 \item [Fig. 4:]  Numerical solution
 (dot) of the Regge trajectory $\orr$
  as function of strong coupling constant $\as$. The solid line
is the numerical fitting given by eqs. (56, 57).
 \item [Fig. 5:]  Normalization factor $n(\as)$ from $\as =0.121$
 to $0.382$
\end{description}


\newpage
\begin{thebibliography}{99}
\baselineskip=14.5pt plus 1pt minus .25pt
\bibitem{bd} B.~Badelek, K.~Charchula, M.~Krawczyk and J.~Kwiecinski,
Rev. of Mod. Phys. 64 (1992) 927
\bibitem{mu} A.H.~Mueller and H.~Navelet, Nucl. Phys. B282 (1987) 727
\bibitem{ta} W.-K.~Tang, Phys. Lett. B278 (1992) 363
\bibitem{kw} J.~Kwiecinski, A.D.~Martin and P.J.~Sutton, Phys. Lett.
B287 (1992) 254 and Phys. Rev. D46 (1992) 921
\bibitem{ba} J.~Bartels,
A.~De~Roeck and M.~Loewe, Z. Phys. C54 (1992) 635
\bibitem{bb} J.~Bartels, M.~Besancon, A.~De~Roeck and J.~Kurzhoefer,
DESY Preprint (1992)
\bibitem{tam} A.H.~Mueller and W.-K.~Tang, Phys. Lett. B284 (1992) 123
\bibitem{ki} R.~Kirschner, Z. Phys. C31(1986) 135 and references therein
\bibitem{li} R.~Kirschner and L.N.~Lipatov, Sov. Phys. JETP 56 (1982) 266
and Nucl. Phys. B213 (1983) 122
\bibitem{ha} H.~Bateman, Higher
Transcendental Functions V.1 (McGraw-Hill,
1953)
\bibitem{ma} W.~Magnus, F.~Oberhettinger and R.D.~Soni, Formulas and
Theorems for the Special Functions of Mathematical Physics,
3rd edition (Springer-Verlag 1966)
%\bibitem{come} In preparation
\end{thebibliography}
\end{document}